\input jnl
\centerline {The Case for Neutrino Oscillations\footnote{$^1$}{Talk 
delivered at the Workshop at the LAMPF neutrino facility, June 1981.}}
\centerline {P. Ramond\footnote{$^2$}{Supported in part by DoE Contract DSR80136je7.}}
\centerline {Physics Department, University of Florida}
\centerline {Gainesville, FL 32611}
\bigskip

The building of a machine capable of producing an intense,
well-calibrated, beam of muon neutrinos is at present regarded 
by particle physicists with keen  interest because of its ability 
of studying neutrino oscillations.

The possibility of neutrino oscillations has long been recognized, 
but it was not made necessary on theoretical or experimental grounds;
one knew that oscillations could be avoided if neutrinos were
massless, and this was easily done by the conservation of lepton
number. The idea of grand unification has led physicists to 
question the existence (at higher energies) of global conservation
laws. The prime examples being baryon number 
conservation which prevents proton decay and lepton number conservation 
which keeps neutrinos massless and therefore free of oscillations.  
The detection of proton decay and neutrino oscillations would
 therefore be an indirect indication of the idea of Grand Unification, and 
therefore of paramount importance.

Neutrino oscillations occur when neutrinos acquire mass in such a way 
that the neutrino mass eigenstates do not match the (neutrino)
eigenstates produced by the weak interactions.  In the following we shall study
the ways in which neutrinos can get mass, first at the level of the 
standard $SU_2 \times U_1$ model, then at the level of its Grand 
Unification generalization.

We start by discussing neutrinos in the standard model.  The
left-handed 
electron-(muon or tau) neutrino is best described in terms of a
two-component 
left-handed (Weyl) spinor, $\nu_L$, which represents a left-handed
particle 
and its right-handed antiparticle, thus conserving CP in first
approximation.  
This is to be contrasted with a charged particle (such as the
electron) 
which is described by two such fields, $e_L$ and $e_R$, conserving C 
and P separately.  The left-handed fields $\nu_{eL}$ and $e_L$ form 
a weak isodoublet $(I_{\rm w} = 1/2)$ and $\nu_L$ has $I_{\rm w}^3 
= + 1/2$.  The standard model interactions involving neutrinos are 
of the form $\nu_L^+ e_L^{},~\nu_L^+ \nu_L^{}$, and $\nu_L^+ e_R^{}$.  
Hence if we assign lepton number $L=1$ to all the fields, these 
interactions conserve $L$.  Note that the electron mass term 
$e_L^+ e_R^{} + c.c$ conserves $L$ as well, and it violates 
weak isospin by $\Delta I_{\rm w} = 1/2$, since $e_R$ is a weak 
singlet.  The left-handed neutrino field can have a mass, the
so-called Majorana mass of the form $\nu_L^T \sigma_2\nu_L^{}$ (in the Weyl 
representation); it violates weak isospin as $\Delta I_{\rm w} = 1$.  
However this Majorana mass clearly violates lepton number $L$ by two 
units.  Hence, no neutrino Majorana mass can develop in a theory with 
$L$-conservation.  In the standard model, the Higgs particle is taken 
to be a weak doublet with $L=0$, which couples the right-handed 
electron field to the weak doublet $(\nu_L, e_L)$.  When it acquires 
a vacuum expectation value, it gives the electron its (Dirac) mass 
and gives the famous relation
$${M_{\rm w}\over M_{\rm z} cos\theta_{\rm w}} = 1\eqno(1)$$
relating the Weinberg angle to the W- and Z-boson masses.  The 
standard model conserves $L$ and neutrinos cannot acquire masses.  
However one can easily generalize it in order to get a massive 
neutrino by breaking $L$ explicitely or spontaneously in the
Lagrangian.  The easiest is to add a Higgs field which is a weak 
isotriplet $(\Delta I_{\rm w} = 1)$ and has $L=-2$.  
The extra Yukawa coupling would then be of the form
$$\left(\nu_L^T~ e_L^T\right) \tau_2\vec\tau ~~
\left(\matrix{ \nu_L\cr e_L\cr}\right)\cdot\vec\phi\eqno(2)$$
where $\vec\phi = (\phi^0, \phi^+, \phi^{++})$ is complex 
in order to preserve electric charge.  If the field $\phi^0$ 
gets a vev, it is known to be very small in comparison to 
that of the Higgs doublet since the relation (1) is experimentally 
good to 3-5\%.  However it generates a Majorana mass for the 
neutrino.  An interesting signature of this coupling would be 
the appearance of a double charged (exotic) Higgs particle in 
the $e^+ e^+$ channel.  However this is only one of many ways 
to obtain massive neutrinos in the standard model.  For instance 
the introduction of explicit $L$-violating terms in the Hamiltonian 
will liberate the neutrino mass and induce it sooner or later in 
perturbation theory.  (Remember that in the standard model
$L$-conservation is the \underbar {only} symmetry that prevents a 
Majorana neutrino mass.)  Hence to go further one has to blend in 
extra theoretical prejudices.  We use those of Grand-Unification 
which loosely speaking says that at some scale one cannot tell a 
quark from a lepton, which means that there exist vector (gauge) 
particles which cause transition between leptons and quarks.  
Assign baryon number $B=1/3~(-1/3)$ for quark (antiquark) and 
$B=0$ for lepton; also set $L=0$ for quarks and antiquarks.  
Thus a vector boson which mediates lepton-antiquark transition 
has $B=1/3$ and $L=1$ and is a color triplet.  Another vector 
boson color triplet changes an antiquark into a quark and has 
$B=-2/3,~L=0$.  In the simplest Grand Unified theory these two 
vector bosons are the same, thus violating $B$ and $L$ separately.  
However since these two have the same value of $B-L = -2/3,~B-L$ 
is conserved, and since the neutrino Majorana mass has $B-L = -2$, 
neutrinos are still massless in the simplest Grand Unified Model, 
although it allows for proton decay, as it well known.

Grand Unified models beyond $SU_5$ introduce fermions not found 
in the standard model and these fermions pave the way for $B-L$ 
violation.  In fact a characteristic of all models beyond $SU_5,$ 
such as $SO_{10},~E_6$ is their extra neutral fermions.

In the following, without showing any particular model, we will 
analyze in terms of the standard model what happens in five different 
types of generalizations for the neutral lepton content of the theory.

The first type of generalization involves more of the usual neutrinos; 
the mass matrix is now purely $\Delta_{\rm w} = 1$:
$$\left(\nu_L^{1\over 2}\right)\left(\Delta I_{\rm w}=1\right)
\left(\nu_L^{1\over 2}\right)\ ,\eqno(3)$$
where $\nu_L^{1\over 2}$ stands for the normal neutrinos (3 in 
the standard model).  Then, as discussed earlier, these neutrinos 
can be made massless by imposing $L$-conservation.

In the second case we have an extra neutrino with $I_w^3 = - 1/2$ 
such as would appear in a theory with $V+A$ currents.  Then the 
most general mass neutrino in the neutral section looks like
$$\left(\nu_L^{1\over 2}~~N_L^{-{1\over 2}}\right) 
\left(\matrix{&\cr \Delta I_{\rm w}=1&\Delta I_{\rm w} = 0,1\cr 
-----&-----\cr \Delta I_{\rm w} = 0,1& \Delta I_{\rm w} = 1\cr}\right) 
\left(\matrix{&\cr \nu_L^{1\over 2}\cr N_L^{-{1\over 2}}\cr}\right)\eqno(4)$$
In the above, $N_L^{-{1\over 2}}$ for the new type neutrino with 
$I_{\rm w}^3 = - 1/2$.  The off-diagonal elements contain the 
so-called Dirac mass and the diagonal elements are the Majorana 
masses.  Since the mass matrix contains a $\Delta I_{\rm w} = 0$ 
component, it has to be understood why it is of the same order of 
magnitude as the $\Delta I_{\rm w} = 1$ component.  Since the matrix 
has no zero eigenvalues, the neutrinos are naturally massive.

The third type of generalization involves adding neutral leptons 
which are mute $(I_{\rm w} = 0)$ under weak interactions.  We denote 
them by $N_L^0$.  The neutral lepton mass matrix now looks like
$$\left(\nu_L^{1\over 4}~~N_L^0\right) \left(\matrix{&\cr 
\Delta I_{\rm w} = 1&\Delta I_{\rm w} = {1\over 2}\cr -----&-----\cr 
\Delta I_{\rm w} = {1\over 2}& \Delta I_{\rm w} = 0\cr}\right) 
\left(\matrix{~~&\cr \nu_L^{1\over 2}\cr N_L^0\cr}\right)\eqno(5)$$
Note the appearance of $\Delta I _{\rm w} = 1/2$ entries in this 
mass matrix.  Barring any global conservation laws, then entries 
will be of the order of the charged leptons and quark masses, say 
$\sim 1$ GeV.  Hence the resulting Majorana mass for the garden 
variety neutrino will be unacceptably large!

The way out is to give the $\Delta I_{\rm w} = 0$ entry a 
\underbar {very} large value $M$.  The mass matrix will then look like
$$\left(\matrix{0&m\cr m&M\cr}\right)\ ,\eqno(6)$$
and will have a small eigenvalue
$$m_\nu \sim m {m\over M}\ ,\eqno(7)$$
i.e. depressed from the usual mass by the ratio of the 
$\Delta I_{\rm w} = 1/2$ to $\Delta I_{\rm w}=0$ scales.  
In Grand Unified theories, such as $SO_{10}$, the $\Delta I_{\rm w} =
0$ scale is $\sim 10^{15}$ GeV, yielding the requisite suppression.

The fourth kind of generalization involves both types of weakly 
interacting extra neutrinos $N_L^{1\over 2}$ and $N_L^{-{1\over 2}}$ 
(a self-conjugate fermion).  The mass matrix now looks like
$$\left(\nu_L^{1\over 2}~~N_L^{1\over 2}~~N_L^{-{1\over 2}}\right) 
\left(\matrix{\Delta I_{\rm w} = 1&\Delta I_{\rm w} =0,1\cr 
\Delta I_{\rm w} = 0,1&\Delta I_{\rm w}=1\cr}\right) 
\left(\matrix{\nu_L^{1\over 2}\cr N_L^{1\over 2}\cr N_L^{-{1\over 2}}\cr}\right)\eqno(8)$$
In the absence of the $\Delta I_{\rm w} = 1$ component, the new 
matrix becomes
$$\left(\matrix{0&0&A\cr 0&0&B\cr A&B&0\cr}\right)\ ,\eqno(9)$$
which, upon diagonalization, gives a massless left-handed neutrino 
and a massive Dirac neutral lepton of mass of the order of the 
$\Delta I_{\rm w} = 0$ mixing.

Lastly one can have a combination of the last two cases, such as 
in the Grand Unified Theory based on $E_6$.  In the above we have 
not included generalizations to neutral fermions with 
$I_{\rm w} = 1,3/2,...$ assignments since they would involve exotic 
charge assignments for their (weak) partners.

Thus when we have in addition to the usual neutrinos a self-conjugate 
fermion (i.e. like $N_L^{1\over 2}$ and $N_L^{-{1\over 2}})$ it is 
more natural to preserve the masslessness of the neutrino; on the 
other hand when the extra fermions are non self-conjugate (i.e. an 
odd number of extra fermions) it becomes rather difficult to preserve neutrino masslessness.

Can we now offer some guesses as to the numerical value of neutrino 
masses and mixing angles?  In general, after diagonalization of the 
charged and neutral lepton mass matrices, the charged current density will look like
$$\left(e_L^+~\mu_L^+~\tau_L^+\right)~U T\left(\matrix{\nu_{RL}\cr 
\nu_{\mu L}\cr \nu_{\tau L}\cr} \right) + ...\eqno(10)$$
where $U$ is a unitary 3x3 matrix coming from the diagonalization of 
the charged lepton mass matrix, and $T$ is a 3x3 matrix (not 
necessarily unitary) obtained by diagonalizing the neutral lepton 
mass matrix.  (The unwritten part of the density (10) involves 
transitions to other particles.)  If we take the ansatz between 
mass and mixing angles
$$\tan^2\theta_{ij} \sim {m_i\over m_j}\ , \eqno(11)$$
and
$${m_e\over m_\mu} \sim {1\over 200};~{m_\mu\over m_\tau}~{1\over 20}\ ,\eqno(12)$$
we see that the $U$ matrix does not mix appreciably the electron into 
the other two leptons, and provides a Cabibbo-like mixing between
$\mu$ 
and $\tau$.  The form of $T$ is much less definite since we do not 
know any neutrino masses.  So we take an example based on $SO_{10}$ 
(the third case discussed above).  The neutral mass matrix is
$$\left(\matrix{M^1& M^{1\over 2}\cr M^{1\over 2}& M^0\cr}\right)\ ,\eqno(13)$$
where $M^{\Delta I_{\rm w}}$ are 3x3 matrices (for 3 families).  Set 
the strengths for $M^{\Delta I_{\rm w}}$ as follows
$$\eqalign{\Delta I_{\rm w} = 0~ \sim~ &m_{\rm x}\cr
\Delta I_{\rm w} = -1/2~ \sim~ &m_{\rm w} \equiv \epsilon m_{\rm x}\ ,\cr
\Delta I_{\rm w} = 1~~~~ &\epsilon^2 m_{\rm x}\cr}\eqno(14)$$
where $\epsilon$ is the hierarchy parameter.  We rewrite the matrix (13) as
$$\left(\matrix{\epsilon^2\hat M^1&\epsilon \hat M^{1\over 2}\cr 
\epsilon \hat M^{1\over 2}& \hat M^0\cr}\right)\ ,\eqno(15)$$
where all $M$ are of the same order.  Then the neutral fermion mass 
matrix is given by
$$\hat M^1 + \hat M^{{1\over 2}T}~{1\over \hat M^0}~\hat M^{1\over 2} 
= T^T DT\ ,\eqno(16)$$
where $T$ is the matrix appearing in (10) and $D$ is a diagonal matrix 
with the neutrino masses as entries.  The point of this exercise is to 
note that the physically relevant parameters (mixing parameters in
$T$, 
mass parameters in $D$) are determined from the knowledge of $\hat
M^1,
~ \hat M^{1\over 2}$ and $\hat M^0$.  Now $\hat M^{1\over 2}$ can, 
under some general assumptions, be related to the charge 2/3 mass 
matrix (this happens in $SO_{10}$), but $\hat M^1$ and $\hat M^0$ are 
not directly related to known physics.  In some schemes (where $\hat 
M^0$ is a perturbation on the Grand Unified scale) it can be argued 
that $\hat M^1$ can be neglected in (16), but this still leaves the 
matrix $\hat M^0$.  So, life is very complicated.  Still one can make 
educated guesses based on specific Grand Unified models.  One obtains, 
more often than not, a \underbar{very} light $\nu_e$ and much heavier 
but comparable $\nu_\mu$ and $\nu_\tau$:
$${m_{\nu_e}\over m_{\nu_\mu}}~\sim~10^{-6}\ ;~~{m_{\nu_\mu}\over 
m_{\nu_\tau}} \ltwid 1\ .\eqno(17)$$
Furthermore one finds very little mixing between $\nu_e$ and $\nu_\mu$ 
or $\nu_\tau$, but large mixing between $\nu_\mu$ and $\nu_\tau$.  
None of these results are ironclad, but they seem to be easier to 
obtain, using the greatest na\"ivet\'e.  Hence, they seem to indicate 
that $\nu_e - \nu_\mu$ oscillations will be all but impossible to 
detect while $\nu_\mu - \nu_\tau$ oscillations would be more apparent.

Now with a ``low energy'' machine such results indicate that one 
should first look for the extinction of the $\nu$ beam, and then 
later for $\nu_\mu - \nu_e$ oscillations.  Moreover these are just 
theories which are not directly coordinated with known phenomenology, 
and it is impossible to gauge their validity.  For the moment one 
would be satisfied with the findings of $\nu$-oscillations
irrespective 
of which way they occur.  This would reinforce our theoretical beliefs 
that global conservation laws are not fundamental and as such would be 
as important as the discovery of proton decay.

\centerline {\bf Acknowledgment}

I wish to thank Profs. F. Boehm and G. Stevenson for asking me to 
participate in the stimulating workshop.  I also wish to thank the 
Aspen Center for Physics where the above was written.

\end